\documentclass[fleqn,twoside]{article}
\usepackage{espcrc2}
\usepackage{epsfig}
\newcommand{\AmS}{{\protect\the\textfont2
  A\kern-.1667em\lower.5ex\hbox{M}\kern-.125emS}}

\newcommand{\ba}{\begin{eqnarray}}
\newcommand{\ea}{\end{eqnarray}}
\newcommand{\bas}{\begin{eqnarray*}}
\newcommand{\eas}{\end{eqnarray*}}
\newcommand{\be}{\begin{equation}}
\newcommand{\ee}{\end{equation}}
\newcommand{\bes}{\begin{equation*}}
\newcommand{\ees}{\end{equation*}}
\newcommand{\bi}{\begin{itemize}}
\newcommand{\ei}{\end{itemize}}

\newcommand{\eqn}[1]{eqn.~\ref{#1}}
\newcommand{\fig}[1]{figure~\ref{#1}}
\newcommand{\tab}[1]{table~\ref{#1}}

\newcommand{\bcentre}{\begin{center}}
\newcommand{\ecentre}{\end{center}}

\font\tenmsb=msbm10 scaled\magstep1
\font\sevenmsb=msbm7 scaled\magstep1
\font\fivemsb=msbm5 scaled\magstep1
\newfam\msbfam
\textfont\msbfam=\tenmsb
\scriptfont\msbfam=\sevenmsb
\scriptscriptfont\msbfam=\fivemsb

\title{
Spectroscopy using the Anisotropic Clover Action}

\author{LHP Collaboration: R.G.~Edwards\address[jlab]{Jefferson
Laboratory, MS 12H2, 12000 Jefferson Avenue, Newport News, VA 23606,
USA}, Urs Heller\address{CSIT, Florida State University, Tallahassee, FL 32306-4120, USA}, David Richards\addressmark[jlab]}

\begin{document}

\begin{abstract}
The calculation of the light-hadron spectrum in the quenched
approximation to QCD using an anisotropic clover fermion action is
presented.  The tuning of the parameters of the action is discussed,
using the pion and $\rho$ dispersion relation.  The adoption of an
anisotropic lattice provides clear advantages in the determination of
the baryonic resonances, and in particular that of the so-called Roper
resonance, the lightest radial excitation of the nucleon.
\end{abstract}

\maketitle

The calculation of the properties of excited states of hadrons
composed of light constituents is complicated by two factors.
Firstly, lattice interpolating operators can be constructed to
transform under the irreducible representations of the cubic group of
the lattice, but then the corresponding excited states belong to many
irreducible representations of the full continuum rotation group.
Secondly, the signal-to-noise ratio for such correlators degrades
rapidly at increasing temporal separations.  The first problem can be
addressed by measuring a matrix of correlators, and identifying
particular states common to the lattice irreducible representations in
the approach to the continuum limit.  The second problem can be
addressed through the use of an anisotropic lattice, having a much
smaller lattice spacing in the temporal direction allowing the
behaviour of the correlators at small separations to be examined over
many more time slices.  In this poster, we will investigate the second
technique.

We use an anisotropic version of the standard (unimproved) Wilson
action which can be constructed from the standard isotropic action by
a simple rescaling of the fields.  In addition to the usual coupling
$\beta$, there is a tunable parameter $\xi_0$, the bare anisotropy,
which we tune to give the required renormalised anistropy $\xi$ where
$\xi = a_s/a_t$, the ratio of lattice spacings in the spatial and
temporal directions~\cite{klassenpure}.

We use the anisotropic version of the usual clover-fermion
action\cite{klassen2,chen} with Dirac operator:
\begin{eqnarray}
\lefteqn{{\cal M} = m_0 - \nu_0 W_0 - \frac{\nu}{\xi_0} \sum_k W_k
  -}\nonumber \\
& & \frac{1}{2} \left[ \omega_0 \sum_{k} \sigma_{0k} F_{0k} +
\frac{\omega}{\xi_0} \sum_{k < l} \sigma_{kl} F_{kl}\right],
\end{eqnarray}
where
\begin{eqnarray}
\lefteqn{W_{\mu} = \frac{1}{2} \left[(1 - \gamma_{\mu}) U_{\mu}(x)
\delta_{x+\hat{\mu},y} + \right.}\nonumber\\
& & \left. ( 1 + \gamma_{\mu} ) U_{\mu}^{\dagger}(x -
\hat{\mu}) \delta_{x - \mu,y} \right].
\end{eqnarray}
This form of the action preserves the projection property of the
Wilson action for $r = 1$.  There are two tunable parameters for the
unimproved Wilson fermion action: $\kappa$, and
either $\nu$ or $\nu_0$.  We will set $\nu_0 \equiv 1$, and then tune
$\nu$.  The clover term introduces a further two
parameters, $\omega$ and $\omega_0$.

In the case of free fermions, where the clover term vanishes,
imposition of the continuum dispersion relation requires
\begin{equation}
\nu^2 =  e^{a_t m} \frac{\sinh a_t m}{a_t m} - r \xi \sinh a_t m.
\label{eq:class1}
\end{equation}
To determine the coefficient of the clover term, we now impose a constant
background field, and the dispersion relations then require
\begin{eqnarray}
 \omega_0 & = & \frac{1}{\xi} \left( \frac{e^{a_t m}}{a_t m} -
 \frac{\nu}{\sinh a_t m}\right) \label{eq:class2}\\
\omega & = & \frac{1}{\xi} \left( \frac{e^{a_t m}}{a_t m} -
 \frac{\nu^2}{\sinh a_t m} \right) \label{eq:class3}
\end{eqnarray}

To proceed beyond the classical level, we take the bare anisotropy
$\xi_0$ from ref.~\cite{klassenpure} to give the required renormalised
anisotropy $\xi$.  We then determine the fermion parameters $(\kappa,
\nu)$ to give the required pion mass, and so that the pion and $\rho$
have the correct dispersion relations.  There remains the question of
what to use for the clover coefficients, and here we impose the
classical constraint of \eqn{eq:class1} to obtain a value of $m$,
whence we obtain $\omega_0$ from \eqn{eq:class2}.  There are two
solutions to \eqn{eq:class1}; for the case of $\nu \longrightarrow 1$,
the correct solution for light-quark physics corresponds to $m = 0$,
and this is the solution we use at our non-zero light-quark
masses. The remaining clover coefficient $\omega$ is given by
\eqn{eq:class3}, and both the $\omega$ and $\omega_0$ are then tadpole
improved using the mean values of the spatial and temporal links.

The calculation is performed on $24^3\times 64$ lattices at $\beta =
6.1$ with a renormalised anisotropy $\xi = 3$; the spatial lattice spacing
$a_s(r_0)^{-1} = 2.04~{\rm GeV}$ giving a spatial lattice
size $L \sim 2.3~{\rm fm}$~\cite{klassenpure}.  
%In addition, elements
%of the calculation were repeated on a smaller $16^3 \times 64$
%lattice, with a spatial extent of $1.7~{\rm fm}$.

\begin{table}
\caption{Fermion parameters for $\nu = 0.810$.}
\label{tab:fermion_params} 
\begin{tabular}{cccc}
 $\kappa$ & Configurations & $m_{\pi} a_t$ & $m_{\rho} a_t $\\ \hline
0.2650 & 67 & 0.137(2)& 0.186(2)\\
0.2660 & 159 &  0.108(1) &  0.167(1)\\
0.2665 & 167 &  0.091(1) & 0.158(2) \\
\hline
\end{tabular} 
\vspace{-0.5cm}
\end{table}
\begin{figure}[t]
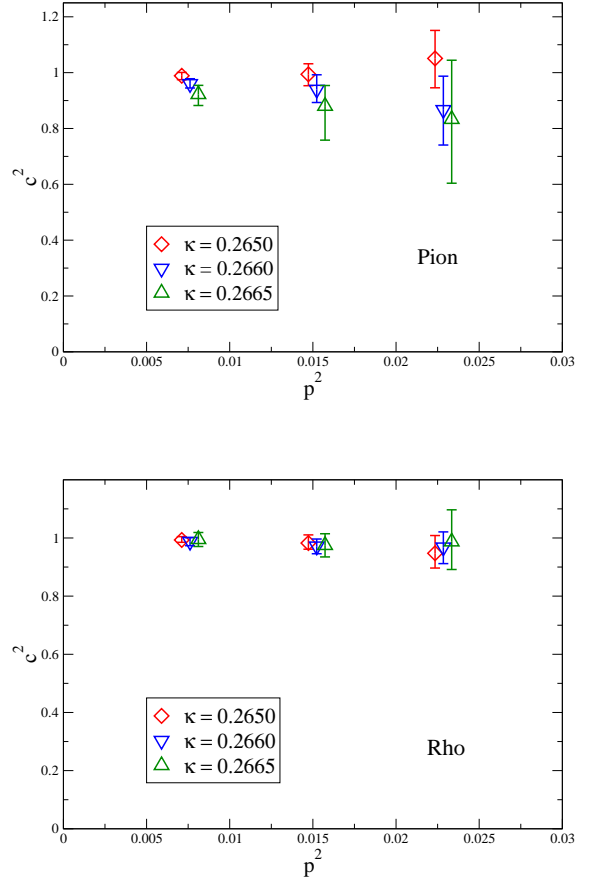

%\vspace{-1.5cm}
\epsfig{width=7.5cm,file=dispersion_chan0.eps}\\[1cm]
\epsfig{width=7.5cm, file=dispersion_chan1.eps}
\caption{The speed of light for the pion and the $\rho$
vs.\ the $\mid \vec{p} \mid^2$ in units of the spatial lattice
spacing.}
\vspace{-0.5cm}
\label{fig:dispersion}
\end{figure}
We choose a range of quark masses corresponding to pion masses between
around $500~{\rm MeV}$ and $900~{\rm MeV}$, and a range of values of
$\nu$ around 0.80.  The determination of the $\kappa$ renormalisation
$\nu$ proved straightforward at heavy quark masses, but more
problematical at light quark mass.  For our final analysis of the
spectrum, we choose $\nu = 0.810$ and the remaning parameters are
listed in \tab{tab:fermion_params}.  The extent to which the
dispersion relations are satisfied for both the pion and the $\rho$ is
shown in \fig{fig:dispersion}.  While the dispersion relation is
clearly well satisfied for the $\rho$ meson, this is less true for the
pion where the tuning of the dispersion relation is increasingly
delicate at lighter quark masses; a possible explanation might be that
at the lightest mass $a_t m_{\pi} \simeq a_t \mid \vec{p} \mid$.

\begin{figure}[tp]
%\vspace{-1cm}
\epsfig{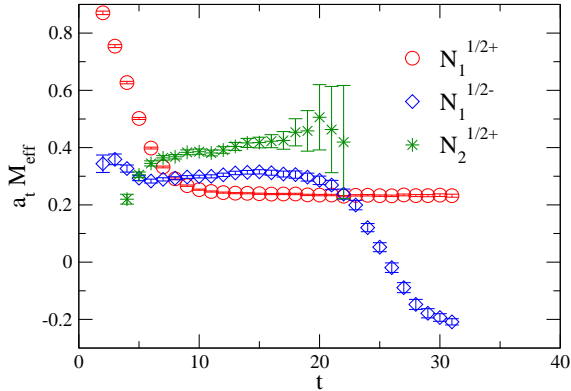}
\caption{The effective masses for the
  correlators corresponding to the nucleon (circles), its parity
  partner (diamonds) and the lower positive-parity excitation
  (crosses) obtained on the anisotropic lattices at $\kappa =
  0.2660$.}
\vspace{-0.5cm}
\label{fig:eff_mass}
\end{figure}

To illustrate the efficacy of anisotropic lattices in the calculation
of the excited hadron spectrum, we will consider the extraction of the
masses of nucleon resonances, and in particular of the nucleon, of its
parity partner, and of the first radial positive-parity excitation; the
latter is of particular interest as the observed light Roper resonance
$N(1440)$ is hard to incorporate within standard quark models.
Propagators are computed from both local and smeared sources, and
fits performed to the smeared-source/local-sink correlators for the
usual nucleon operator $N_1$, and for the ``bad'' nucleon operator
$N_2$ which couples upper and lower quark components and is
conventionally assumed to overlap the first excited nucleon state.
The effective masses for these correlators
are shown in \fig{fig:eff_mass}.  A clear plateau is exhibited
for each of these channels, and this persists to the lightest
pseudoscalar mass studied.

\begin{figure}[t]
%\vspace{-1cm}
\epsfig{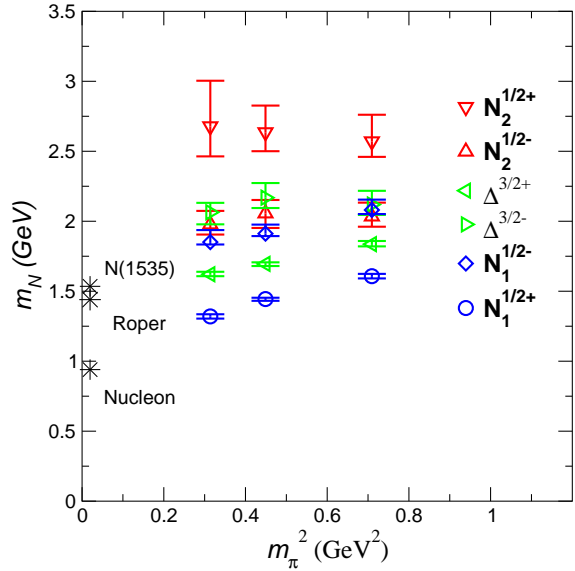}
\caption{The masses of some
  lowest-lying nucleon resonances, using
  $r_0$ to set the scale.}
\vspace{-0.5cm}
\label{fig:spectrum}
\end{figure}
A reliable calculation of the masses of the lowest lying nucleon
resonances of both parities can be made, something not readily
possible with the isotropic clover action, and the measured spectrum
is shown in \fig{fig:spectrum}.  The
calculation reveals that, for the quark masses probed in this study,
the ordering of the masses of the states is $N^{*1/2+} > N^{1/2-} >
N^{1/2+}$ as expected in a simple quark model; furthermore we find
reasonable consistency between the masses of the lightest
negative-parity state determined using the two interpolating operators
$N_1$ and $N_2$.  The large mass of the first positive-parity
excitation was originally suggested using the matrix-correlator
method~\cite{ukqcd}, but there is evidence that it may not persist to
light-quark masses~\cite{frank}.

In this work we have demonstrated the efficacy of using an anistropic
lattice for the calculation of the masses of nucleon resonances.
Whilst the tuning of the action in the heavy quark sector is
straightforward, the tuning is more problematic at light quark masses.
A more extensive analysis will include the use of the full planoply of
techniques aimed at delineating the spectrum, notably the measurement
of a matrix of correlators and the use of Bayesian statistics both to
extract higher radial excitations and to effectively use data close to
the temporal source.  Further work will include the determination of
the light-quark hybrid meson spectrum.

This work was supported in part by DOE contract DE-AC05-84ER40150
under which the Southeastern Universities Research Association (SURA)
operates the Thomas Jefferson National Accelerator Facility,
and by DE-FG02-97ER41022.


\begin{thebibliography}{1}
\bibitem{klassenpure} T.~Klassen, Nucl.\ Phys. B533, 557 (1998).
\bibitem{klassen2} T.R.~Klassen, Nucl.\ Phys. B (Proc.\ Suppl.) 72, 918 (1999).
\bibitem{chen} P.~Chen, Phys.\ Rev.\ D64, 034509 (2001).
\bibitem{ukqcdneg}
M. Gockeler {\it et~al.}, Phys.~Lett.~B {\bf 532}, 63  (2002).
\bibitem{ukqcd} C.~Allton \textit{et al.} (UKQCD Collaboration),
  Phys.\ Rev.\ D47, 5128 (1993).
\bibitem{frank} F.X.~Lee \textit{et al.}, these proceedings, hep-lat/0208070.
\end{thebibliography}
\end{document}